\documentclass[sigconf,authorversion,nonacm]{acmart}

\AtBeginDocument{%
  }

\usepackage{graphicx} % Required for inserting images
\usepackage{xcolor}
\usepackage{hyperref}
\usepackage{natbib}% Citation support using natbib.sty
\usepackage{amsfonts}
\bibpunct[, ]{(}{)}{;}{a}{}{,}% Citation support using natbib.sty
\usepackage{amsmath}
\graphicspath{{./figures/}}

\usepackage{soul} % for strikethrough
\newcommand{\stkout}[1]{\ifmmode\text{\sout{\ensuremath{#1}}}\else\sout{#1}\fi} % make the strikeout work for equations

\date{October 27, 2023}

\begin{document} %{

\title{Fifth Generation IMC: Expanding the scope to Profit, People, and the Planet}


\author{Stewart Pearson}
\email{stewart.pearson@consilient-group.com}
\author{Edward Malthouse}
\orcid{0000-0001-7077-0172}
\email{ecm@northwestern.edu}
\affiliation{%
  \institution{Medill School, Northwestern University}
  \streetaddress{1845 Sheridan Road}
  \city{Evanston}
  \state{Il}
  \country{USA}
}

\renewcommand{\shortauthors}{Pearson and Malthouse}

\begin{abstract}
This editorial outlines an expanded scope for the next (fifth) generation of integrated marketing
communication. It identifies key market forces that gave rise to this evolution and describes a
trajectory of where Integrated Marketing Communication (IMC) has been and where it is going.
The central shift is moving from primarily
focusing on one stakeholder to multiple ones, including people (employees and society), the planet
(environment), and profits. It identifies examples from industry that exemplify multi-stakeholder
decision-making and uses the examples to suggest research questions that academics and practitioners
should address. Examples and research directions are organized around marketing strategy,
communication media and messages, and measurement systems.
\end{abstract}

\maketitle

%===================================================================
\section{Introduction: State of Marketing and Advertising}

In the last two decades Advertising and Marketing have been transformed by computation. Now they are on the cusp of the next paradigm shift, one that will open expansive new opportunities for research and innovation. Following a trend that started with the 2008 Recession and accelerated with Covid-19 and the extreme climate events of 2020-22, the Environmental, Societal and Governance (ESG) movement emerged with its focus on multiple stakeholders. The emerging world view has resulted in a change in multiple disciplines. The purpose of this editorial is to propose a vision for an updated Integrated Marketing response to these challenges.

In today's hyper-competitive markets, brands no longer differentiate themselves by unique selling propositions, but by meeting diverse values embraced by multiple stakeholders. The previous focus on shareholder returns is giving way to a balanced approach meeting the needs of employees and customers as well as shareholders. And with the ESG-movement enterprises recognize and report environmental and social as well as financial metrics.

This signals a paradigm shift for Advertising and Marketing. Advertising today is mostly automated
and programmatic, driven by technology (adtech) and data. It is served through a fragmented media
ecosystem and managed for short-term behavioral metrics. Targeted and personalized advertising
resulted in media and audience fragmentation. Mass markets dissolved into niche audiences and
``taste communities'' \citep{andjelic2020business}.

Advertising was once described as ``truth well told'' and ``the art of persuasion,'' (Levenson 1987). Now all functions of Marketing have similarly become technology-led (martech) with most communication and commerce delivered over the five monopolistic digital platforms (Apple, Alphabet, Amazon, Meta, and Microsoft) aided by thousands of software tools. Marketing once built enduring brand franchises with widespread appeal. Now it mostly delivers sales into niche audiences for increasingly ephemeral brands.

A further consequence is the decline of mass media that once served a national audience, and the local media that once connected local communities. On digital platforms news content is algorithmically targeted with recommender systems based on past content consumption, reinforcing ingrained beliefs and behaviors. The negative outcomes from data-driven digital advertising are now societal and wide: a deepening lack of trust in the media and institutions, and a widening polarization of people and communities.

Over its history advertising has never been well regarded. Delivered programmatically through digital platforms, advertising's reputation further declined. The surveillance business model, based on personal data collected and traded with lax regulation, is increasingly resisted by consumers (ad blocking), and is under investigation by governments around the world.

New business models offer alternatives to advertising. Netflix, the leading streaming television
network in the US, built its business through subscription rather than advertising revenue. Local
news media, long supported by advertising, are also shifting their revenue models towards
subscriptions \citep{kim2020DJ}. With news and entertainment behind paywalls, NYU Professor Scott Galloway calls advertising a ``tax on the poor.''

Yet advertising executed by knowledgeable marketers created and maintained long-existing brands like
those of Procter \& Gamble, Coca-Cola, and Philip Morris (Marlboro), generating positive economic impact. In recent decades, before the programmatic era, marketing insights and advertising creativity fueled the wide and enduring global appeal of brands like Apple, Nike, and Louis Vuitton. The advertising that built these brands was generative, not reinforcing. The messaging was suffused with insight, wit, humor, entertainment, and narrative.

In 2022, as reported by Kantar, of the world's ten most valuable enterprises rated by their brand
contribution, no fewer than eight are technology or payment platforms. Digital marketing and
advertising no longer build enduring brands. As \citet{srnicek2017platform} summarized: ``Data extraction became the key method of building a monopolistic platform and siphoning off revenue from advertisers.''

%===================================================================
\section{The New Emerging Paradigm}
In the first half of the 20th century advertising and marketing developed with the growth of the
mass market and economy. Marketing generated the consumer demand that met the supply of the
products that enterprises manufactured and distributed. Advertising's contribution was summarized
by one of its leading practitioners, David Ogilvy who promised ``We Sell or Else''
\citep{ogilvy2023ogilvy}

In the second half of the century, the importance and scope of advertising and marketing accelerated, led by the U.S. and U.K. From the 70's onwards Anglo-Saxon neoliberal economics superseded the Keynesian economics of the 50's and 60's. The Keynesians emphasized the role of government in managing demand and maintaining a level of social equality. The neoliberals prioritized total shareholder value more than anything else, including the needs of employees and customers.

A new world view is now emerging. It is changing multiple disciplines including philosophy, science, economics, sociology, and psychology. We summarize these changes here, before interrogating the response of the advertising and marketing disciplines. 

Philosophers are reconsidering the scope of markets. \citet{sandel2012money} wrote how American
society has been changing from a market economy to what he calls a "market society, a way of life
in which market values and market reasoning reach into every sphere of life, including education,
health care and military service.'' He argues for overcoming ``an allergy we have $\cdots$ to bringing
ethics, morality and virtue into public discourse.''

Scientists are calling for action on climate change with increasing urgency. The 2022 report of the UN's Intergovernmental Panel on Climate Change (IPCC) forecasts that to avoid very dangerous warming, carbon emissions need to peak within three years, and fall rapidly after that. This will require urgent changes to all aspects of the design, manufacturing, marketing, and distribution of goods.

Economists are proposing a reformation in their discipline. \citet{Raworth2017doughnut}' Doughnut Economics model calls for operating in ``a safe and just place for humanity'', between a social foundation and an ecological ceiling. Enterprises should be responsive to the social well-being of employees (after decades of stagnant earnings) and to the impact of their products on society (after rising morbidities). In evaluating their environmental impact, enterprises should be responsive not only to investors but to all citizens experiencing environmental degradation. 

Sociologists and psychologists are reporting the negative effects of two decades of social media
usage, noting how the platforms are majority funded by advertising. NYU professor and social
psychologist Jonathan Haidt \citep{haidt21} demonstrates the case that social media is a major
cause of harm to teenage girls. Technologist James Williams, formerly with Google, now teaches
Philosophy at Cambridge and writes ``Digital technologies undermine the human will at both
individual and collective levels'' \citep{williams2018stand}. 

Advertising and marketing must operate in a new paradigm in which:
\begin{itemize}
\item The goals set and metrics reported by enterprises are expanding. They will continue to report their financial measures of revenues and profits, but now they go beyond and report on their Environmental, Social, and Governmental (ESG) impact. We ask how marketing planning changes when tasked with multiple goals and metrics in multiple domains?
\item The diversity of stakeholders served by enterprises is expanding. They continue to prioritize shareholders, and customer-centricity is always important (mission critical in the age of Amazon). There is a growing appreciation of the value of employees and partners and the wider communities of producers with whom they work. The ESG movement compels measuring citizens they impact, locally and globally. We ask how marketing communication changes with multiple stakeholders? 
\item The complexity and connectivity of enterprises are expanding. The penetration of mobile devices, advances in telecommunications, and global supply chains make their operations and marketplaces hyperconnected networks. For marketers this means that the drivers of influence are dispersed, their effects non-linear, and the outcomes unpredictable and often chaotic. We ask how marketing science changes with complexity.
Our contention is that advertising and marketing must develop new theories and new models, creating opportunities for research. In developing the models, and conducting the research, researchers will find it critical to collaborate with peers from the other domains identified above.
\end{itemize}

This is not only a theoretical issue. Advertising and marketing are already transforming and tasked
with communicating environmental and societal commitments. At the 2022 Cannes Lions Festival of
Creativity, 85% of the Grands Prix creative winners went to Cause-related or Social-Purpose
campaigns. For practitioners, whether ESG goals should matter is already a controversial issue
\citep{harrison2021can}. 

In what follows we present the case for a new generation of Integrated Marketing Communications (IMC) as a framework for incorporating the ESG goals.

%===================================================================
\section{IMC: past and future}

The article's title promises a next generation for IMC, which we see as the key concept for planning how advertising and marketing contribute to ESG as well as commercial goals. We briefly review the earlier generations and describe a vision for the future of IMC.

The first generation of IMC was described in \citet{schultz1993new} and prescribed coordinating messages across media channels including mass advertising, sales promotion, direct marketing, and public relations. The book hypothesized that a core benefit from such integration is synergy, where the outcome from two or more coordinated channels is greater than the sum of the parts.

Strategic and tactical coordination would seem to be essential in creating the type of behavioral change required to address ESG challenges, which is partly why we have decided to ground our solution in IMC. ESG goals are often presented as conflicting with financial goals. We contend that in the new paradigm there can be synergy. By integrating marcom across multiple stakeholders, enterprises can deliver long-term brand value growth at the same time as they contribute to the social foundation and meet their commitments to the environment.

The second generation of IMC was proposed in \citet{schultz2003imc}. One of their most important advances was to propose the IMC process, which integrated measurement, financial accountability, and customer data from direct marketing with the power of branding and advertising. Skepticism around ESG goals from the financial community suggests that any solution must be grounded in sound metrics to have credibility. Another prominent aspect of The Next Generation book was customer centricity, the primary focus on the customer stakeholder. As discussed above, we argue that this singular focus must evolve to consider and balance the utility of other stakeholders as well.

In 2004 the potential of social media, multi-way interactions, networked consumers, and the increase of
consumer power was only starting to be realized. Many authors \citep{schultz2012cm, finne2013rethinking}
expanded the scope of IMC to include such customer engagement, and we see this as the third generation
of IMC. With the global reach of social media, we now see people engaging with brands as fans,
collaborators, co-promoters, and citizens motivated by ESG issues.

More recently the divide between traditional or analogue media, in particular television, and digital media, is disappearing. As television became connected and digital, we saw a 4th generation of IMC emerging. Digital video on the platforms, first YouTube and now Tik Tok, is coordinated with Connected TV (CTV) according to IMC practices, blurring the boundaries between many formerly distinct media and channels. CTV began to go mainstream in 2020 at the onset of the pandemic. The integration of connected television and video with all other digital media and content completes the vision of the original proponents of IMC.

The next generation of IMC is well-suited to play a role in advancing ESG because the theory and practice of coordinated messages, measurement, co-creation, and engagement are well-understood. Our thesis is that the scope of IMC can be expanded to these dimensions, to develop a new paradigm for advertising and marketing in integrating multiple stakeholders, multiple metrics, and multiple domains, as we discuss next.

\begin{table*}[htb]
    \centering
    \begin{tabular}{|p{1.0in}|p{2.5in}|p{3.0in}|}
\hline
 & Earlier Generations of Integrated Marketing Comms 
 & Impact Marketing for People, Planet, and Profit \\
\hline
Stakeholders  
  & Customer centricity as principal focus of Marcom 
  & Multiple stakeholders: shareholders, customers, employees, fans, collaborators, and involved citizens\\
Goals \& Metrics 
  & Financial optimization \& ROAS via revenue from sales to customers, and Shareholder Value to investors  
  & Win-win-win outcomes for financial, environmental \& social goals, with strategies realizing
    synergies for stakeholders\\
\hline
Data Sources 
  & Profile data \& behaviors for personalization \& targeting against media \& channel spend \& activity 
  & Expanded through data from domains beyond sales \& marketing systems: economical, \& environmental,
    social \& governmental\\
\hline
    \end{tabular}
    \caption{Integrated Marketing's New Paradigm Summarized}
    \label{tab:IMC}
\end{table*}

We refer to this as Integrated Marketing for Profit, People, and Planet, or IMP3. We need to retool IMC to deal with
the most difficult challenges of the day. The shifts in the paradigm are summarized in Table\ref{tab:IMC}, and the
implications for the practice in Table~\ref{tab:impact}. The sections following the tables discuss in further depth and propose questions for researchers, prompted by emerging practices.

\begin{table*}[htb]
    \centering
    \begin{tabular}{|p{1.2in}|p{2.5in}|p{3.0in}|}
\hline
 & Earlier Generations of Integrated Marketing Comms 
 & Impact Marketing for People, Planet, and Profit\\
\hline
Strategy \& Planning 
  & Integration of Media, Content \& Channel Plans \& Investment to achieve financial outcomes (ROAS)  
  & Integration of Engagement, Content, Media \& Channel Plans across Multiple Stakeholders to achieve
    positive financial, environmental \& societal outcomes\\
\hline
Communications: Media \& Messaging 
  & Persuasive brand messages with promotional offers measured by sales 
  & Truthful \& persuasive brand \& co-created content \& conversations, with intrinsic and extrinsic
    rewards to all stakeholders, measured by engagement and sales \\
\hline
Insights, Metrics \& Measurement 
  & Generated by research \& analysis of consumer behaviors \& sales in market 
  & Integration of measurement systems from other domains: positive psychology (well-being); \&
  environmental \& societal impact\\
\hline
    \end{tabular}
    \caption{Impact Marketing's New Practices Summarized}
    \label{tab:impact}
\end{table*}

Marketing has always been the function within an enterprise charged with insights into the customer, competition, and category. Impact Marketing expands the sources and scope of insight generation. Impact Marketers will access social and environmental data sources beyond the market to monitor, interpret and scenario plan.

%===================================================================
\section{Strategy and Planning for Multiple Stakeholders}

Our concept of IMP3 is distinctive in moving away from the primary focus on customer and advertiser/client to
manage multiple stakeholders and coordinate multiple activities across multiple data points and influences in
multiple domains. Different advertising stakeholders are identified in \citet[Figure 1]{yun20ja} and
\citet{copulskyIJA22}.

The understanding of the importance of multiple stakeholders is now in the mainstream of business and brand strategy. Brands are encountering challenges that illustrate the opportunities for researchers:

\begin{itemize}
\item The founder of the clothing brand Patagonia, Yvon Chouinard, has given his family's shareholding in the brand and declared that the earth is Patagonia's only shareholder. Research topic: How do high-level announcements like Patagonia's impact customer behavior and Patagonia's financial performance?
\item Former Unilever CEO Paul Polman wrote \citep{baddipudi2023net} of his experience: ``I don't
think our fiduciary duty is to put shareholders first. I say the opposite. What we firmly believe
is that if we focus our company on improving the lives of the world's citizens and come up with
genuine sustainable solutions, we are more in synch with consumers and society and ultimately this
will result in good shareholder returns.'' Research topic: How to quantify the impact of
communication strategies of enterprises like Unilever on consumers around the world?
\item Fast growing chocolate brand Tony's Chocolonely has a mission focused on slavery-free chocolate. Central to his company's mission and his primary stakeholders are the African cocoa producers: ``Right now there is slavery on cocoa farms in West Africa. This is a result of the unequally divided cocoa chain. Tony's Chocolonely exists to change that. Illegal child labor and modern slavery are against the law. It needs to stop.'' Research topic: What impact does a brand's commitment to stakeholders like the producers of its product ingredients have on its other stakeholders: its distributors and consumers?
\item Airbnb, Adidas, Apple, Amazon, Coca-Cola, Facebook, Nike, Uber, Warby Parker, and YouTube have all donated to organizations and initiatives fighting inequality and systemic racism or have committed to award grants to black-owned businesses. Research topic: What impact does a purely financial donation to a cause have on a brand's principal stakeholders?
\end{itemize}

%===================================================================
\section{Communications, Media, and Messaging for Multiple Stakeholders}
We also see industry communications that motivate research questions. AB InBev, the largest global beer company, demonstrated multiple stakeholder communications, media, and messaging in the launch Michelob ULTRA Pure Gold, the first organic beer from a national brewer. The brand communicates its distinctiveness with messaging relevant to consumers and to the producers of the organic barley ingredient.

Television and digital advertising reached consumers using celebrity musician endorsements.
One spot features actor and singer Becky G enjoying her night at a local bowling alley, talking about her penchant for gold after being served the Michelob ULTRA Pure Gold light lager by her server. Another features singer songwriter Sugaray Rayford enjoying a beach bonfire with friends, celebrating the enjoyment of the beer with company.

The most distinctive aspect of the Michelob ULTRA Pure Gold marketing is an initiative ``Contract for Change'' that supports American barley farmers that want to transition conventional fields to try organic production for the first time. The brand, alongside agronomists, provides financial assistance to farmers as they navigate the costly and time- consuming steps required to grow certified organic crops. Pure Gold will purchase transitional barley at a premium price, supporting farmers during the transition window.

Allbirds illustrates a company, launched in 2016, with a business model designed from the outset to serve multiple stakeholders, again consumers and suppliers. The proposition to consumers is captured in the names of its products such as ``Tree Breezes'' and ``Wool Runners'' shoes. It is supported by the partnership it operates to offer lightly used previously owned shoes. The proposition to its tight supply chain is demonstrated by the investments and commitments it makes in ``Regenerative Agriculture,'' ``Renewable Materials,'' and ``Responsible Energy.''

These examples suggest the research topic: What synergies are created when multiple stakeholders see coordinated messaging and what impact does a purely financial donation to a cause have on a brand's principal stakeholders?

The scales of the challenge for established enterprises to expand their stakeholders are illustrated by cases where communications to multiple stakeholders are not coordinated. Starbucks built its brand through establishing a friendly and accessible ``third place'' in consumers' lives. Starbucks is a master of IMC, increasing CLV through innovative media and channels including mobile rewards and commerce. Starbucks announced ``a bold, multi-decade aspiration'' to cut carbon, water, and waste each by 50 percent by 2030 and give back to the planet by becoming resource positive in the future. Yet even as it responded on the environment, Starbucks found itself with a new challenge on another front, in contesting employee demand for union representation. Research topic: how do positive communications to meet one stakeholder goal (here on the environment) interact with negative communications on another (here on social equity), and what is their impact on a third (the consumers generating financial performance)?

McDonalds offered Happy Meals for adults.
Walker\footnote{\href{https://kotaku.com/mcdonalds-adult-happy-meal-cactus-plant-flea-market-1849619811?utm_source=morning-10&utm_medium=email&utm_campaign=20221006&utm_content=article9-readmore}{McDonald's Workers Are Begging
People To Stop Ordering Adult Happy Meals}} reports that employees felt overburdened by the promotion and posted messages to social media begging customers not to order the meals. It seems that management made the decision to run the promotion without considering the effects on employee stakeholders, as well as the waste created by the plastic toys. Research topic: how do brands coordinate their communications to employees, when consumer-facing staff are responsible for delivery to the customer.

To address a lack of research and physician knowledge about skin conditions for those with darker skin, Vaseline created a website to help dark-skinned people diagnose and treat skin conditions. Research topic: how to measure the effects of this initiative both in terms of financial benefit to Vaseline, and to other stakeholders such as lives saved and reduced healthcare costs to governments and insurance companies?

%===================================================================
\section{Measuring Value Creation for Multiple Stakeholders}
Having the right managerial metrics in place will be critical. If there are no metrics in place to measure ESG outcomes, then ESG activities may not be valued by the organization.

Direct marketing, customer relationship management and advertising has focused on customer lifetime
value (CLV), the discounted sum of expected future cash flows with a relationship objective of for
decades \citep{kumar2016conceptualizing}. It represents the expected profit from a relationship and is used to allocate marketing resources and measure effectiveness. It does not, however, consider other stakeholders.

Similarly, marketers have attempted to measure the contribution of brand equity (BE)---the added
value that the thoughts, feelings, and beliefs about a brand give to a product or service
\citep{farquhar1989marketing}---to total corporate financial value, led by Interbrand and Kantar, now independent, formerly part of WPP. Kantar's BrandZ and WPP's Brand Asset Valuator (BAV) measures brands on their salience, esteem, meaning, and uniqueness in the perception of consumers.

While CLV and BE will remain important, marketers must expand the set of metrics and consider tradeoffs between the competing objectives of different stakeholders. Further, there is a need to develop quantitative measures of the contributions made to environmental impact and societal well-being. We already see hints of this shift happening in industry, although it is at an early stage of development. For example, Consultancy Brand Finance measures BE on traditional consumer-focused metrics (Consideration, Familiarity, Preference, Recommendation). They are adding other dimensions including an employee Score (based on auditing) and a reputational score (based on investment analyst recommendations). The prospects are exciting for the scope and value of IMP3.

By measuring their effects on the lives of people (beyond their purchasing behavior) brands can play a greater and positive role in their lives. We contend that by better understanding people's lives, and by expanding Marketing's ambition to integrate vital environmental and societal communications and measure their impact, the resulting synergies will make Marketing more powerful.

%===================================================================
\section{Multiple Stakeholders and the Case for Synergy}
Earlier iterations of IMC planning and models generally focused on a single, short-term objective,
such as estimating the probability of a click or a purchase. For example, news organizations made
money by selling advertising and retaining subscribers. Increasing the number of ads on a page will
increase ad revenue in the short term (more exposures to sell), but too many ads may cause
disengagement in the future (e.g., not coming back to the site or installing ad blockers) and may
cause subscribers to churn in the future and reduce retention rates and CLV
\citep{miroglio2018effect}.

Models need to consider these complicated interactions and long-term (unintended) consequences. Planning for multiple stakeholder outcomes increases the complexity, but the experience of IMC is that synergy is a significant contributor to all outcomes. This simple exposition makes the point.

\begin{figure}
    \centering
    \includegraphics[width=.5\textwidth]{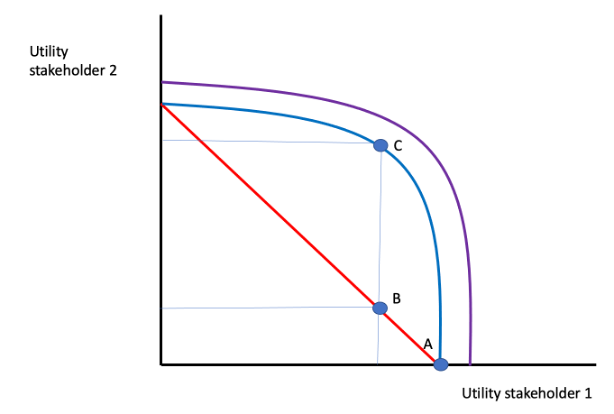}
    \caption{Moving from a zero-sum game to win-win-win}
    \label{fig:pareto}
\end{figure}

Figure~\ref{fig:pareto} shows an example, plotting the utility derived by two stakeholders, e.g., customers and employees. Each line shows different ways of allocating resources to the two stakeholders. The red 45-degree line shows a zero-sum game, where any increase in utility for Stakeholder 2 must be offset with an equal decrease in utility for Stakeholder 1, e.g., moving from point A to B in the figure. But what if the red line does not reflect the nature of the tradeoffs, and the blue line is a better description of reality?

\citet{malthouse19RS} study the situation for two competing objectives, selecting items to maximize customer utility versus selecting items to maximize ad revenue from sponsored recommendations. They find a concave curve that bows outward. If the blue line is true then small reductions in utility for one stakeholder (e.g., point A to point C) increase the utility substantially for the other stakeholder. Once all the factors are considered, point C could be more desirable than point A. Another possibility is that by generating higher utility for both stakeholders, the curve shifts outward over time, e.g., the purple curve. In such a case both stakeholders could be substantially better off. In the case of customer and employee stakeholders, such a shift could occur, for example, through well-designed government incentives or regulations.

In the long run the winners will be brands and marketers who understand how everything they do affects their multiple stakeholders. It is why in game theory, the winners in the long run are those who collaborate. We believe that prioritizing one stakeholder group and excluding others is short-sighted and doomed to failure.

The evidence that coordinating and amplifying a consistent message drives value across channels.
According to
Kantar,footnote{\href{https://www.prnewswire.com/news-releases/more-than-half-of-marketers-miss-opportunity-to-boost-brand-effectiveness-by-57-by-not-getting-their-multichannel-campaigns-right-says-kantar-millward-brown-300582918.html}{More
than half of marketers miss opportunity to boost brand effectiveness by 57% by not getting their
multichannel campaigns right}} integrated campaigns are 31 percent more effective at building
brands. Now the challenge is to research how integrating ESG communications with multiple
stakeholders is even more effective in building brands as well as meeting ESG commitments.

%===================================================================
\section{Conclusion}
We have described a vision, $\mathrm{IMP}^3$, for how marketing must evolve to address the evolving needs in the marketplace with an expanded focus on multiple stakeholders and considerations. We have identified examples from industry that allow us to glimpse this future approach to marketing and pointed up possible avenues for research.

%===================================================================
%\bibliographystyle{tfcad}
\bibliographystyle{ACM-Reference-Format}
\bibliography{refs.bib}

\end{document}